\documentclass[reprint,amssymb,amsmath,superscriptaddress,twocolumn,showkeys,aps,prx]{revtex4-1}   %{revtex4-1}
\usepackage{graphicx}
\usepackage{amssymb}
\usepackage{amsmath}
\usepackage{dcolumn}
\usepackage{bm}
\usepackage{hyperref}

\begin{document}

%%%%%%%%%%%%%%%   TITLE AND AUTHORS  %%%%%%%%%%%%%%%

\title{Dynamics of a single-atom electron pump}
\keywords{Single-electron pump, single atom, silicon, nano-electronics, quantum metrology}

\author{J. van der Heijden}
\email{j.vanderheijden@unsw.edu.au}
\author{G. C. Tettamanzi}
\email{g.tettamanzi@unsw.edu.au}
\author{S. Rogge}
\affiliation{School of Physics and Australian Centre of Excellence for Quantum Computation and Communication Technology, UNSW, Sydney, Australia}

%%%%%%%%%%%%%%%   ABSTRACT  %%%%%%%%%%%%%%%
\begin{abstract}
Single-electron pumps based on isolated impurity atoms have recently been experimentally demonstrated. In these devices the Coulomb potential of an atom creates a localised electron state with a large charging energy and considerable orbital level spacings, enabling robust charge capturing processes. In these single-atom pumps, the confinement potential is hardly affected by the periodic driving of the system. This is in contrast to the often used gate-defined quantum dot pumps, for which a strongly time-dependent potential leads to significantly different charge pumping processes. Here we describe the behaviour and the performance of an atomic, single parameter, electron pump. This is done by considering the loading, isolating and unloading of one electron at the time, on a phosphorous atom embedded in a silicon double gate transistor. The most important feature of the atom pump is its very isolated ground state, which can be populated through the fast loading of much higher lying excited states and a subsequent fast relaxation proces. This leads to a substantial increase in pumping accuracy, and is opposed to the adverse role of excited states as observed for quantum dot pumps due to non-adiabatic excitations. The pumping performances are investigated as a function of dopant position, revealing a pumping behaviour robust against the expected variability in atomic position.
\end{abstract}
%%%%%%%%%%%%%%%%%%%%%%%%%%%%%%%%%%%%%

\maketitle

\section{Introduction}

Quantized charge pumps can be used as single electron sources~\cite{Kouwenhoven:1991aa, Pothier:1992aa, Blumenthal:2007aa, Giblin:2012aa, Rossi:2014aa, Tettamanzi:2014aa}, which have potential applications in electron quantum optics~\cite{Ubbelohde:2015aa, Kataoka:2016aa} and can be used as a new standard for the Ampere in quantum metrology~\cite{Milton:2010aa, Pekola:2013aa} when they can be operated with a sufficiently high accuracy~\cite{Keller:1996aa, Giblin:2012aa, Jehl:2013aa, Rossi:2014aa, Stein:2015aa}. Essential to semiconductor electron pumps is to control the energy states of a quantum dot~\cite{Hohls:2012aa, Giblin:2012aa, Rossi:2014aa, Stein:2015aa}, an impurity atom~\cite{Lansbergen:2012aa, Roche:2013aa, Tettamanzi:2014aa} or both~\cite{Yamahata:2014aa, Wenz:2016aa}, in order to capture, isolate and emit a fixed number of electrons in every cycle of the electron pump. The most common charge pumps are based on quantum dots where the shape of the dot strongly depends on the AC driving voltage of the pump. Due to the drastic reshaping of the confinement potential for the electron during the pumping cycle, this type of charge pump is named the dynamical quantum dot pump. This dynamic behaviour of the electron confinement is in sharp contrast with electron pumps based on single dopant atoms, where the electron confinement originates from the fixed 1/r Coulomb potential of the atom. This leads to a significantly different pumping process and requires a new model to accurately describe the single-atom charge pumps.

Recent improvements in electron confinement~\cite{Kaestner:2009aa, Rossi:2014aa}, pulse shapes~\cite{Giblin:2012aa} and readout techniques~\cite{Yamahata:2011aa, Fricke:2013aa} have led to highly accurate single-electron pumps based on dynamical quantum dots. The pumping process of these pumps has been studied extensively~\cite{Kaestner:2015aa} and is accurately described by the universal cascade decay model~\cite{Kashcheyevs:2010aa}. The best performing pumps have achieved an uncertainty below 0.2 ppm at a driving frequency around 500 MHz~\cite{Stein:2015aa}. Here back-tunnelling processes, where the electron tunnels back to its source, are found to be the main source of systemetic errors at these high pumping frequencies~\cite{Kashcheyevs:2010aa, Fricke:2013aa}. At even higher frequencies non-adiabatic excitations~\cite{Liu:1993aa, Kataoka:2011aa} start to play a detrimental role, as the fast AC driving voltage produces unwanted excitations in the quantum dot, which can substantially increase the back-tunnelling rate and thereby decrease the accuracy of the electron pump. 

Dopant atom electron pumps, using the fixed atomic potential as confinement, have the capability to lower these systematic errors. The Coulomb potential of a donor atom provides a highly localized electronic state, which isolates the electron during the pumping process and thereby reduces the chances of back-tunnelling. Furthermore, the large energy spacing between the ground and excited states of an atom lowers the chance of non-adiabtic excitations. An inevitable consequence of the fixed atomic confinement and isolated electronic state is the possibility of an error in the loading process of the electron to the ground state of the atom. However, as our model explains, an accurate pumping cycle is achieved due to the loading via excited states. This is in sharp contrast with the behaviour of dynamical quantum dots, where the accurate description based on the universal decay cascade model ascribes the pumping error to the ratio between the back-tunnelling probabilities of N and N+1 electrons, leaving out any loading and unloading errors~\cite{Kashcheyevs:2010aa}. 

To accurately describe the behaviour of a single-atom electron pump, we present a model that takes into account the loading, back-tunnelling and unloading processes as the three main stages of the pumping cycle, and capture all possible sources of systematic errors. It is found that the most accurate charge pumping is achieved only when multiple energy states of the atom contribute to the loading stage of the pump cycle. In this scenario the tunnel rate from the leads to the ground state of the atom is insufficient to load an electron. Instead an electron tunnels from the source to one of the excited states of the atom with greater efficiency, because of the higher tunnel rates associated with the spatially more extended excited states. After this initial step, the electron relaxes to the ground state, where, due to the slow tunnel rate between the ground state and the source, back-tunnelling is substantially suppressed. This beneficial effect of the excited states for the single-atom pump clearly opposes their adverse role for dynamical quantum dot pumps. 

An incomplete relaxation from the atomic excited states to the ground state is found to be the main source of error in the pumping process via atoms. This is further investigated by studying the influence of the pumping frequency and relaxation rates on the pumping accuracy. The last part of this work focusses on the influence of the donor position in this scheme and shows that the behaviour of the single-atom electron pump is robust for displacements of the donor over a distance of 30 nm.

\section{Single-atom electron pump model}

In the following section a complete description of the single-atom electron pump model is given, by outlining the geometry of the pump, specifying the used postulates and providing the theoretical framework.

\subsection{Device description}

\begin{figure}[t]
\begin{center}
\includegraphics[width=86mm]{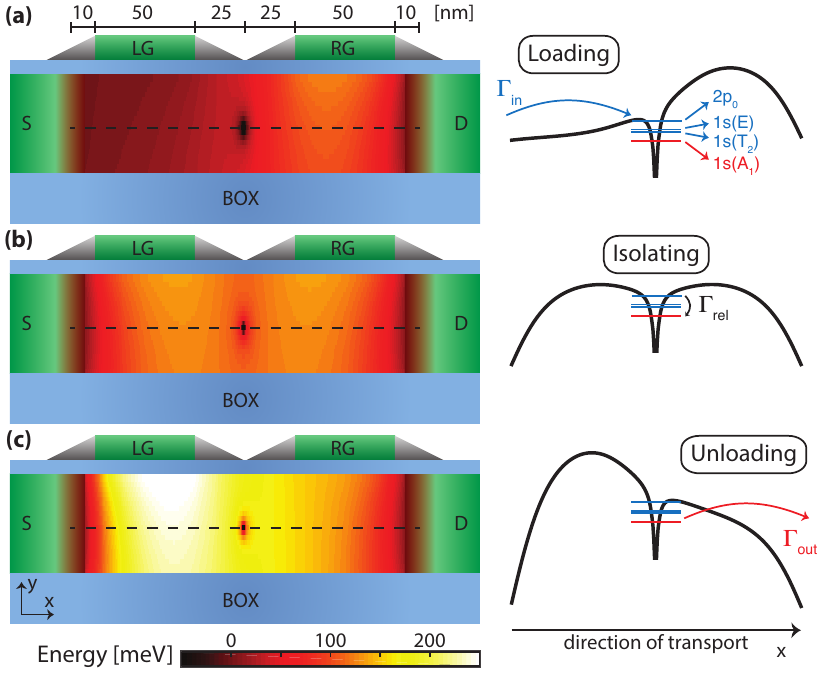}
\caption{(a,b,c) Calculation of the potential profile in a cut through the middle, i.e. in the direction of the width, of the double gate transistor as used in Ref.~\cite{Tettamanzi:2014aa}, where one dopant atom is placed in the centre of the transistor channel with dimensions 170-50-20 nm (length-width-height). The voltages used to create these profiles identify the (a) loading (b) isolating and (c) unloading stages in the pumping cycle. A plot of the profile in the x (length) direction indicated with a dashed line is shown on the right side for each of these profiles and the relevant tunnel rates are indicated.}
\label{Figure1}
\end{center}
\end{figure}

The model is directly compared to experiments on a single-atom electron pump. The specific geometry of this device, which has recently demonstrated single electron pumping~\cite{Tettamanzi:2014aa} is used to calculate the tunnel rates, involved in pumping the electrons one-by-one through the transistor, for our model. As shown in Fig.~\ref{Figure1}, both the left gate (LG) and right gate (RG) are 50 nm long and are separated by 50 nm. The source and drain are separated from the gates by silicon nitride spacers to get to a total channel length of 200 nm, however some diffusion of the source and drain dopants under the spacers is taken into account, which reduces the effective length to 170 nm. In our model, the width and height of the channel are taken to be 50 and 20 nm respectively and the dopant potential is initially placed in the centre of the transistor channel in all directions, i.e. length, width and height. The effect of changing the position of this potential along the length direction is discussed in the last section. For the experiments, the single-atom electron pump is mounted on a cold finger and measured at 4.2 K. A low noise battery operated measurement setup was used to measure the source/drain current and to apply dc voltages to the gates. Typical measurements were acquired at $V_{SD}\sim$ 0 mV and by applying a sinusoidal rf input to the left gate via the bias-tee in a typical single parameter configuration~\cite{Kaestner:2015aa, Tettamanzi:2014aa}.

%%%%%%%%%%%%%%%   FIGURE 1  %%%%%%%%%%%%%%%
%%%%%%%%%%%%%%%%%%%%%%%%%%%%%%%%%%%%%

\subsection{Model postulates}

The energy spectrum of the phosphorous atom is, in our model, always assumed to be equal to the known bulk values~\cite{Jagannath:1981aa} and fixed throughout the pumping cycle. Especially at a distance of 10 nm from the $Si/SiO_2$ interface (i.e. middle in the height direction, see black dashed lines in Fig.~\ref{Figure1}), this assumption is justified, as the excited state energies and charging energy are expected to approach the bulk values \cite{Rahman:2009aa, Calderon:2010aa}, which has been experimentally confirmed~\cite{Fuechsle:2012aa, Roche:2012aa}. Furthermore,  we find a maximum electric field during the unloading stage of the pumping cycle of 4 MV/m, where in the more important loading stage the field never exceeds 3.5 MV/m. For these electric fields it has been shown that the excited state spectrum of a donor atom does not greatly change~\cite{Lansbergen:2008aa} and also the charging energy of the atom is stable~\cite{Rahman:2011aa}. The singlet $1s(A_1)$ state with a binding energy of 45.6 meV is used as the ground state in our model together with the lowest three excited states, being the valley-orbit triplet $1s(T_2)$ and doublet $1s(E)$ states seperated by 11.7 and 13 meV from the ground state respectively and the $2p_0$ state seperated by 34.1 meV. The binding energy of the second electron is taken to be 2 meV~\cite{Burger:1984aa}.

Contrary to the decay cascade model, our model considers only the tunnelling via the first electron state, i.e. the $D^0$ state, which restricts the voltages applied to the gates to values for which the energy level of a second donor-bound electron stays above the source and drain Fermi energy during the entire pump-cycle. The doubly occupied, negatively charged donor state is referred to as $D^-$. Although this restriction in gate voltage space denies the modelling of the pumping current regions that are related to more than one electron and likely does not include the complete voltage space where one electron is pumped, it still allows the capturing of the unique physics of a single-atom pump. This because the large charging energy of the donor atom allows the capturing of the first electron at a much lower energy and therefore much lower voltages than the second electron. Lastly, for single-impurity charge pumps, and in agreement with our theoretical approach, only the 1 electron plateau can be effectively used to estimate the accuracy of the pump~\cite{Yamahata:2014aa}

\subsection{Theoretical framework}

The model presented here is based on three consecutive calculation steps. First the potential profile in the channel is calculated by solving the Poisson equation, see Fig.~\ref{Figure1}. Second, this potential profile is used to estimate the tunnel rates between the different energy levels of the donor atom and the source and drain leads, see Fig.~\ref{Figure2}(b), by using the Wentzel-Kramers-Brillouin (WKB) method, as in Eq.~\ref{Eq1}. And third, these tunnel rates are used in a rate-equation model, as in Eq.~\ref{Eq2} to Eq.~\ref{Eq4}, to find the occupation probabilities of all the states during the pumping cycle, see Fig.~\ref{Figure2}(c), which consequently gives the average amount of electrons transferred per cycle, as in Eq.~\ref{Eq5}.

The Poisson equation is solved by using the finite difference method for different combinations of voltages on the left and right gate, while keeping the source and drain leads grounded, consistent with the experimental situation. This solution gives the three dimensional potential landscape in the transistor for each set of gate voltages, see Fig.~\ref{Figure1}. Subsequently the tunnel rates between the energy levels of the atom and the source and drain leads are calculated with the use of the one dimensional WKB method:
\begin{align}
\label{Eq1}
\Gamma_{i}(t) = \xi \Gamma_0 \int_{0}^{x_{dop}}e^{-2x\sqrt{U(x,t)-E_{i}(t)}\sqrt{\frac{2m^*}{\hbar^2}}}dx
\end{align}
Here $\Gamma_{i}(t)$ is the tunnel rate to or from energy state $i$ at time $t$ of the pumping cycle, $U(x,t)$ is the potential in the transistor at position $x$, as defined in Fig.~\ref{Figure1}, and at time $t$. $E_{i}$ is the energy of the considered donor state and $m^*$ is the effective mass in silicon. The integral is taken from the end of either the source or drain lead, denoted by position $0$, to the position of the dopant $x_{dop}$. Furthermore, $\xi$ denotes the degeneracy of the state and $\Gamma_0$ is the attempt frequency, which symbolises the maximum tunnel rate in this system and is assumed to be 100 THz \cite{note1}.

The behaviour of the single-atom electron pump is described by the occupation probabilities of all the atomic states during the pump cycle. The estimated tunnel rates are used in a rate equation model to calculate the occupation probabilities of all the atomic energy states. The occupation probabilities are represented by the vector $P(t) = \left[\begin{array}{ccccc}P_u(t) & P_0(t) & P_1(t) & P_2(t) & P_3(t)\end{array}\right]$, consisting of the probability for the atom to be unoccupied followed by the probabilities of one electron to be in the $1s(A_1)$, $1s(T_2)$, $1s(E)$,  and $2p_0$ states at time t. The occupation probabilities during the pumping cycle are found by solving the following set of equations:
\begin{align}
\label{Eq2}
\sum\limits_{i=u,0..4} P_i(t) = 1
\end{align}
\begin{align}
\label{Eq3}
 dP/dt = M(t) \times P(t)
\end{align}
\begin{align}
\label{Eq4}
M = \begin{pmatrix} -\sum\limits_{n=0..3} \Gamma^i_{n} & \Gamma^o_{0} & \Gamma^o_{1} & \Gamma^o_{2} & \Gamma^o_{3} \\ \Gamma^i_{0} & -\Gamma^o_{0} & \Gamma^r_{1} & \Gamma^r_{2} & \Gamma^r_{3} \\ \Gamma^i_{1} & 0 & -\Gamma^o_{1}-\Gamma^r_{1} & 0 & 0 \\ \Gamma^i_{2} & 0 & 0 & -\Gamma^o_{2}-\Gamma^r_{2} & 0 \\ \Gamma^i_{3} & 0 & 0 & 0 & -\Gamma^o_{3}-\Gamma^r_{3} \end{pmatrix}
\end{align}
Here $\Gamma^i_n$ is the in-tunnel rate (i.e. from the source or drain to the atom state $n$), $\Gamma^o_n$ the out-tunnel rate (i.e. from the atom state $n$ to the source and drain) and $\Gamma^r_m$ the relaxation rate of the donor excited state $m$ to the $1s(A_1)$ ground state. The tunnel rates, as calculated by Eq.~\ref{Eq1} with taking spin and valley degeneracies into account for the tunnel-in rates, are regarded as $\Gamma^i$ if the energy of the donor state is below the Fermi level of the source and drain and $\Gamma^o$ otherwise. For the source and drain contacts a Fermi-Dirac distribution of electrons at a temperature of 4.2 Kelvin is assumed. The rate equation is numerically solved using the explicit Runge-Kutta Dormand-Prince method.

The relaxation rates from the excited states to the ground state play an important role for the single-atom electron pump. Estimations of the relaxation rates from the $1s(T_2)$, $1s(E)$ states to the $1s(A_1)$ state are found in the 1 to 100 GHz range~\cite{Hubers:2005aa, Zhukavin:2007aa, Soykal:2011aa}. Relaxation from the $2p_0$ state is expected to follow a double relaxation process, where it first relaxes to either the $1s(T_2)$ or the $1s(E)$ state~\cite{Hubers:2005aa,Zhukavin:2007aa}, with an estimated relaxation rate of several tens of GHz~\cite{Tsyplenkov:2008aa}. The total relaxation rate from the $2p_0$ state is measured around 5 GHz~\cite{Vinh:2008aa}. To incorporate these findings in our model, only direct relaxations to the $1s(A_1)$ are included (see Eq.~\ref{Eq4}), where the relaxation rates are set to 10 GHz for all three excited states. As the pumping accuracy depends strongly on the assumed relaxation rates, the effect of a change in the relaxation rates is investigated later in the paper.

\section{Typical pump cycle}

\begin{figure}[t]
\begin{center}
\includegraphics[width=86mm]{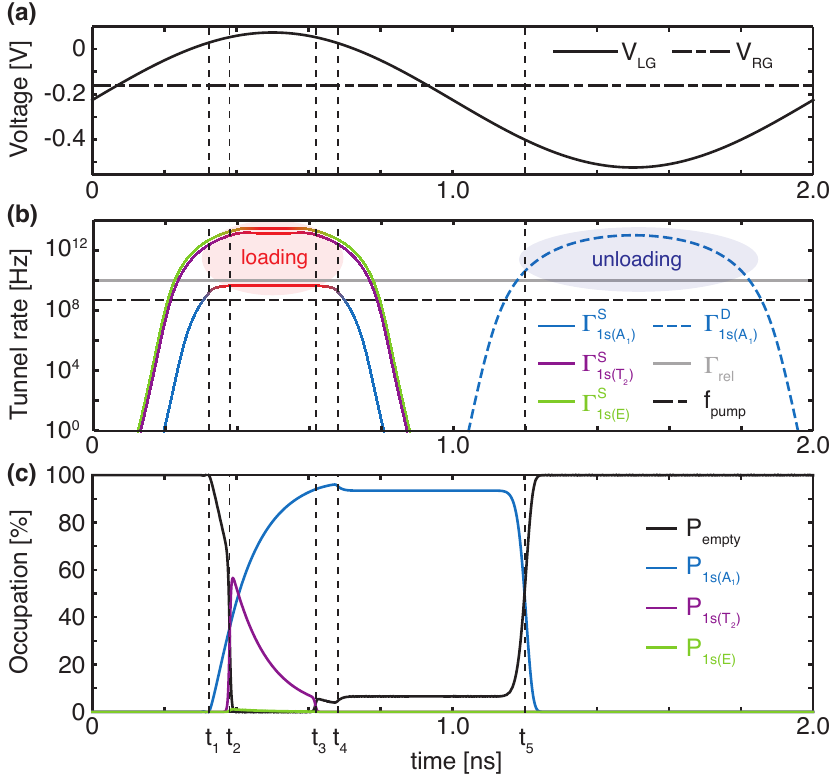}
\caption{Typical operation of the single-atom electron pump. (a) Time dependence of voltages $V_{LG}$ and $V_{RG}$ during the cycle. (b) The relevant tunnel rates from the source (solid lines) and to the drain (dashed line) during the pumping cycle for the $1s(A_1)$ state (blue), the $1s(T_2)$ state (purple) and the $1s(E)$ state (green). When a state is below the Fermi level of the source, the line is colored red. A grey line shows the relaxation rates for the $1s(T_2)$ and $1s(E)$ states to the $1s(A_1)$ state and a black dashed-dotted line shows the pumping frequency. (c) The corresponding occupation probabilities for the relevant single electron states of the atom as a function of time, where five moments in the cycle are highlighted with $t_1$ to $t_5$, as discussed in the text.}
\label{Figure2}
\end{center}
\end{figure}

Next, the operation of the single-atom electron pump will be discussed by examining the time dependent voltages, tunnel rates and occupation probabilities during a typical pump cycle (as displayed in Fig.~\ref{Figure2}). This cycle is simulated with $V_{RG}$ = -165 mV and $V_{LG}$ = -220 mV, while a 300 mV AC voltage is added to the left gate, see Fig.~\ref{Figure2}(a). Note that these are not the optimal conditions for the pump (on average, only 0.9 electrons are pumped per cycle), but it shows a pumping cycle where both the benefits of excited states are visible and possible error mechanisms are displayed. The phase of the driving voltage is chosen such that the cycle starts between the unloading stage (low $V_{LG}$) and loading stage (high $V_{LG}$), where the initial condition is chosen as $P_u(0) = 1$ and $P_{0..3}(0) = 0$. As this initial condition only holds if the unloading stage is complete, it is iteratively updated when the final state differs from the initial state.

%%%%%%%%%%%%%%%   FIGURE 2  %%%%%%%%%%%%%%%
%%%%%%%%%%%%%%%%%%%%%%%%%%%%%%%%%%%%%

In Fig.~\ref{Figure2} the critical moments in the pump cycle of the single-atom electron pump are denoted with $t_i$, where i ranges from 1 to 5, and the dashed lines show the corrosponding tunnel rates (Fig.~\ref{Figure2}(b)) and occupation probabilities (Fig.~\ref{Figure2}(c)). The loading stage starts at times $t_1$ and $t_2$, where the $1s(A_1)$ state starts to slowly load at $t_1$ and the $1s(T_2)$ quickly loads at $t_2$. At $t_2$ the $1s(E)$ state barely contributes to the loading process, as shown in Fig.~\ref{Figure2}(c), which is due to the fact that the $1s(T_2)$ crosses the Fermi energy a short moment before the $1s(E)$ state and fills almost completely within that time, due to the high tunnel-in rate. The back-tunnelling processes from the atomic states start at $t_3$ ($1s(T_2)$) and $t_4$ ($1s(A_1)$). The back-tunnelling error from the $1s(A_1)$ ground state can only be reduced by lowering the tunnel rate from this state to the source. In contrast, the back-tunnelling error from the ($1s(T_2)$) can be reduced by either a faster relaxation rate to the ground state, or a longer time ($\lvert t_3-t_2 \rvert$) to relax to the ground state. After $t_4$ the loading of electrons to the atomic ground state is finished and all other excited states are empty. Importantly, this model indicates that for a single-atom quantum electron pump, a large part of the occupation of the ground state originates from a loading process to the $1s(T_2)$ excited state and a subsequent relaxation, which is a completely different loading mechanism than the loading process of dynamical quantum dot electron pumps, for which this occupation happens mainly via a directly filling of the ground state. Finally, at $t_5$ the unloading of the electron from the ground state to the drain occurs.

%%%%%%%%%%%%%%%   FIGURE 3  %%%%%%%%%%%%%%%
\begin{figure*}
\begin{center}
\includegraphics[width=178mm]{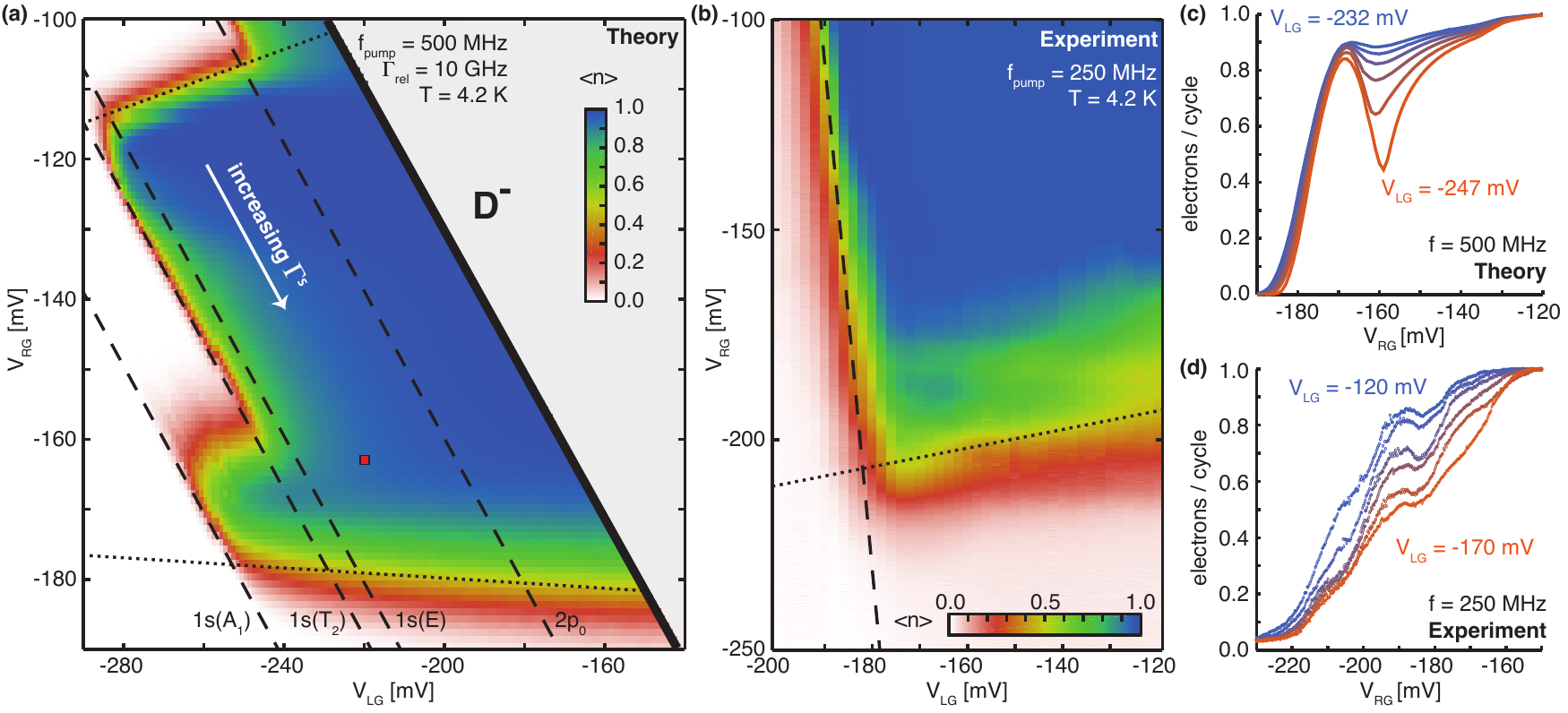}
\caption{The average number of electrons pumped per cycle by this single-atom electron pump. (a) Number of pumped electrons ($\left<n\right>$ in Eq.~\ref{Eq5}) as a function of the DC voltages $V_{LG}$ and $V_{RG}$, using an AC sinusoidal driving voltage on $V_{LG}$ with a 300 mV amplitude and a 500 MHz frequency in the calculations. The red square corresponds to the pumping cycle as shown in Fig.~\ref{Figure2}. The solid, dashed and dotted lines indicate the boundaries of the pumping plateau, as discussed in the text. (b) The model is compared with experimental data from a single-atom electron pump, which is operated with a 250 MHz and $\sim$100 mV driving signal. The left and bottom boundaries found in the model are also found in the experiment. (c) Six line-plots at $V_{LG}$ (-247, -244, -241, -238, -235, -232) showing a stepwise increase of pumped electrons in the model, as a result of the involvement of the excited states in the loading process. (d) Line-plots taken from (b) for voltages $V_{LG}$ (-120, -130, -140, -150,- 160, -170). In this experiment similar steps of the pumping current as found by the model are observed as a function of $V_{RG}$.}
\label{Figure3}
\end{center}
\end{figure*}
%%%%%%%%%%%%%%%%%%%%%%%%%%%%%%%%%%%%%

\section{Comparison between the model and experiment}

The total probability of pumping a single electron from the source to the drain, denoted as $\left<n\right>$, can be estimated by only accounting for the electrons that go to and come from the drain (or source) in Eq.~\ref{Eq3}, which gives:
\begin{align}
\label{Eq5}
d\left< n \right>/dt &= \sum\limits_{n=0..3} \Gamma^{o,d}_n (t) P_n (t) - \sum\limits_{n=0..3} \Gamma^{i,d}_n (t) P_u (t)
\end{align}
Here $\Gamma^{o,d}_n$ and $\Gamma^{i,d}_n$ are the tunnel rates to and from the drain lead. In the visualisation of the data in Fig.~\ref{Figure3} the average number of electrons pumped per cycle $<$$n$$>$, defined as~$\frac{I_{SD}}{ef}$, is shown. Fig.~\ref{Figure3}(a) shows the average number of electrons per cycle as a function of the DC voltages on the left and right gates when a 500 MHz sinusoidal AC voltage with 300 mV in amplitude is added to the left gate, i.e. in a single parameter charge pumping configuration. A plateau of current close to the ideal 1 electron per cycle is found, which is in agreement with experiments~\cite{Tettamanzi:2014aa}. The solid black line in Fig.~\ref{Figure3}(a) shows the boundary of the model where the $D^-$ state equals the source and drain Fermi level at the lowest potential in the pump cycle. The crossings of the one electron ground and excited states with the Fermi level are also shown in Fig.~\ref{Figure3}(a) (dashed lines). A clear increase in pumping accuracy is observed when the loading of the electron into the ground state is possible via an intermediate excited state with subsequent relaxation. This underlines the important role of the excited states in the accuracy of the single-atom pump. 

The model is compared to the experiment on a single-atom electron pump, which is operated with a driving signal of a 250 MHz frequency and around a 100 mV amplitude, see Fig.~\ref{Figure3}(b). The bottom and left boundary indicated by the model are also found in the experimental data, where both the direction and broadening of these boundaries show strong similarities. The increase in pumping current caused by the improved loading of the ground state via the excited states appears as several steps in the model when line cuts are taken at different values of $V_{LG}$ as shown in Fig.~\ref{Figure3}(c). These line cuts are compared with line cuts in the experimental data of Fig.~\ref{Figure3}(b), as shown in Fig.~\ref{Figure3}(d). The similarities between the experiment and theory in the steplike increase of the pumping current towards the plateau of 1 electron per cycle, both in line shape and DC voltages, demonstrates the good agreement between the model and the experiment. Furthermore, these similarities confirm the strong positive influence of the excited states of the atom on the behaviour of this pump geometry as seen in the experiment. Indeed, for single-atom pumps the excited states play a crucial role to establish an effective pumping cycle, greatly contrasting the adverse role of excited states for dynamical quantum dot pumps through non-adiabatic excitations.

\section{Discussion of the pumping accuracy}

The operational region of the single-atom electron pump has, next to the boundaries caused by the crossing of the Fermi energy (dashed and solid lines in Fig.~\ref{Figure3}(a)), two boundaries caused by the limits on the tunnel rate between the atom and the source (dotted lines in Fig.~\ref{Figure3}(a)), as also discussed in Ref.~\cite{Tettamanzi:2014aa}. This tunnel rate increases in the direction of a more negative $V_{RG}$, with a small dependence on $V_{LG}$, see Fig.~\ref{Figure3}(a). First, the boundary found at the bottom of Fig.~\ref{Figure3}(a), around a -180 mV voltage on the right gate and almost independent of $V_{LG}$ is discussed, where the exact gate dependencies depend strongly on the position of the atom. At this boundary the tunnel rate from the $1s(A_1)$ ground state to the source is too fast at the moment this state crosses the Fermi level, leading to electrons tunnelling back to the source as marked with time $t_4$ in Fig.~\ref{Figure2}. This is the same limit as considered in the cascade decay model~\cite{Kashcheyevs:2010aa}. At the other side of the plateau a boundary is found for $V_{RG}$ around -110 mV, which has a small dependence on $V_{LG}$. At this boundary the loading stage of the atom is incomplete, caused by insufficient tunnel rates to the $1s(T_2)$ and $1s(E)$ to get to the full occupation of the atom within the time these states are below the Fermi level of the source. As the time of the pump-cycle that these states spend below the Fermi level depends on $V_{LG}$, this boundary has a dependence on the left gate voltage as well as the right gate voltage. A similar boundary can be seen for the $1s(A_1)$ ground state around -160 mV $V_{RG}$ and will be present for the $2p_0$ state at a higher $V_{RG}$. In the center of the single electron plateau, far away from every boundary and as visible in Fig.~\ref{Figure2}, the accuracy of the pump is limited by the ratio between the pumping frequency and relaxation rates from the excited states. Here the main systemic error comes from the back-tunnelling of electrons from the excited states (shown at $t_3$ in Fig.~\ref{Figure2}). The occupation probability of the excited states, at the moment these states cross the Fermi level at $t_3$, strongly depends on how fast the electron relaxes to the ground state and the time this state spends below the Fermi level ($\lvert t_3-t_2 \rvert$), which depends on the pumping frequency.

%%%%%%%%%%%%%%%   FIGURE 4  %%%%%%%%%%%%%%%
\begin{figure}[t]
\begin{center}
\includegraphics[width=86mm]{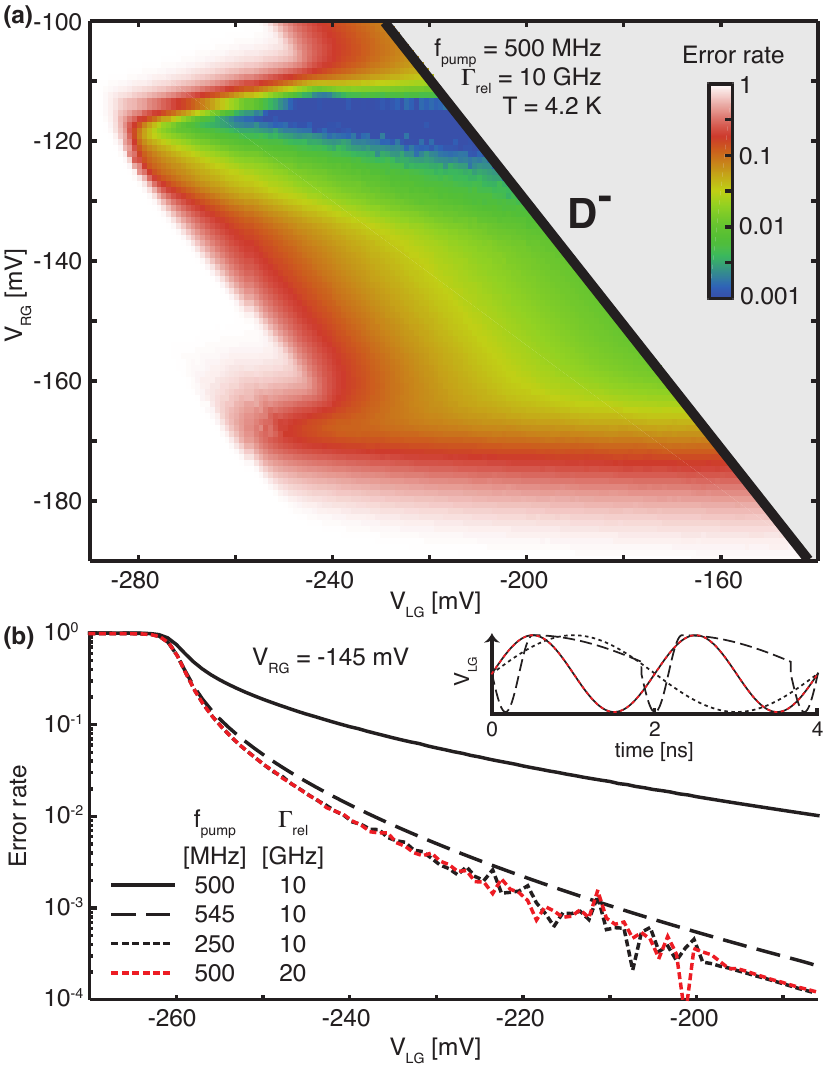}
\caption{The accuracy of the single-atom electron pump. (a) The same result as in Fig.~\ref{Figure3}(a), but now shown as the number of electrons per cycle less than the ideal 1 electron per cycle on a logarithmic scale. (b) Linecut at -145 mV $V_{RG}$ (solid line) compared with a pump cycle with half the pumping frequency (black dotted line). The same improvement in accuracy is found when the relaxation rates are doubled (red dotted line). Furthermore, a more effective pump cycle which uses more time in the loading stage and less in the unloading stage, is investigated (dashed line). The inset shows the voltages during the different cycles.}
\label{Figure4}
\end{center}
\end{figure}
%%%%%%%%%%%%%%%%%%%%%%%%%%%%%%%%%%%%%

In Fig.~\ref{Figure4} the theoretical error rate of the single-atom electron pump is shown, defined as the difference with 1 electron per cycle, which illustrates the current quantisation properties of the electron pump in terms of accuracy. The exact behaviour of the pump's theoretical error as a function of the DC voltages on the left and right gate is shown in Fig.~\ref{Figure4}(a). To investigate the effect of the ratio between the relaxation rate and pumping frequency, a simulation with the same pumping frequency of 500 MHz, but twice as fast relaxation times (20 GHz) is compared to a simulation with the same relaxation time (10 GHz), but at half the pumping frequency (250 MHz). Both give the same result for the accuracy in the center of the plateau (shown in Fig.~\ref{Figure4}(b)), which is two orders of magnitude more accurate than the original pumping cycle. Alternatively, to slow down the pumping frequency in the crucial loading stage, but still get a decent current output by ramping up the frequency in the unloading stage, several other waveforms have been proposed~\cite{Giblin:2012aa, Yamahata:2014aa, Stein:2015aa}. We compare our initial model to the accuracy of a model that has an effective frequency of 150 MHz in the loading stage and 1.5 GHz in the unloading stage to get a total frequency around 545 MHz (dashed line in Fig.~\ref{Figure4}(b)). This model shows an improvement of almost 2 orders of magnitude on the original model, while keeping the pumping frequency and relaxation rate constant, explained by the increased time the excited states stay under the Fermi level of the source~\cite{Giblin:2012aa, Stein:2015aa}. These findings emphasize that the relaxation from excited states to the ground state of the atom is a parameter of major importance for the accuracy of the single-atom electron pump. The understanding of the possible relaxation paths between these valley-orbit excited states and available methods to increase these relaxation rates are essential to improve the accuracy of the single-impurity electron pumps, which have already shown pumping currents at frequencies of a few GHz~\cite{Tettamanzi:2014aa, Yamahata:2014aa}.

%%%%%%%%%%%%%%%   FIGURE 5  %%%%%%%%%%%%%%%
\begin{figure}[t]
\begin{center}
\includegraphics[width=86mm]{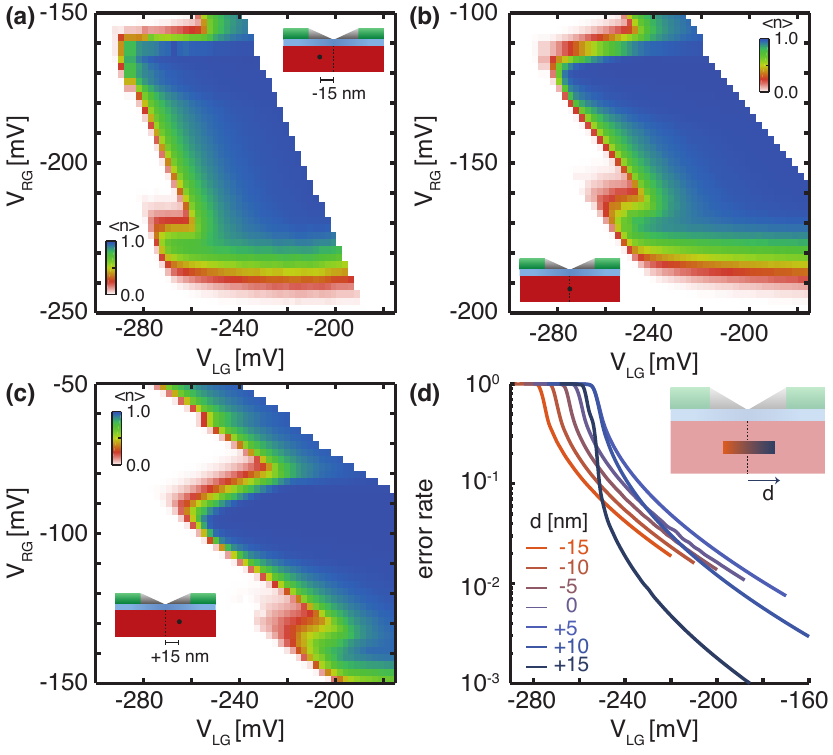}
\caption{Dependence of the single-atom pump behaviour on the placement of the atom. (a) The phosphorous donor is displaced 15 nm towards the left gate from the centre. (b) The donor is in the centre. (c) The donor is displaced 15 nm towards the right gate from the centre. All simulations use a sinusoidal AC voltage on the left gate of 300 mV and 1 GHz and 20 GHz relaxation rates are assumed for all excited states, in order to keep the ratio between the pumping frequency and relaxation time equal to those in Figs.~\ref{Figure2}, \ref{Figure3} and \ref{Figure4}(a), while reaching a faster computational speed. Furthermore, a lower resolution than in Figs.~\ref{Figure3}(a) and \ref{Figure4}(a) is used, to reduce the total computation time. The same color scale as in Fig.~\ref{Figure3}(a) is used. (d) Pumping accuracy as a function of $V_{LG}$ for different donor positions, where the traces are taken at $V_{RG}= (-145 + \gamma d)$ mV to compensate for the shift of the plateau. Here $d$ is the position relative to the center of the transistor and $\gamma$ is 3 mV/nm.} 
\label{Figure5}
\end{center}
\end{figure}
%%%%%%%%%%%%%%%%%%%%%%%%%%%%%%%%%%%%%

\section{Influence of the atom placement}

Finally, to study the feasibility of the single-atom electron pump, we have benchmarked the robustness of this pump as a function of the position of the atom. In Fig.~\ref{Figure5} the pumping behaviour with the atom in the centre (\ref{Figure5}(b)) and moved 15 nm towards both the left (\ref{Figure5}(a)) and right (\ref{Figure5}(c)) gate are calculated, using a pumping frequency of 1 GHz and relaxation rates of 20 GHz. This is a fair comparison with the previous calculations, as this configuration keeps the same ratio between pumping frequency and relaxation rates. As it was shown in the previous sections, this ratio determines the maximum accuracy of the pump and we only expect slight changes to the boundaries of the pumping plateaus as an effect of the higher frequencies. Due to the change in capacitive coupling from the atom to the left and right gates, the slopes of the boundaries of the pumping plateau change with the change of atom position. Using this observation, the difference between the shape of the plateaus found in the theoretical model and the experimental data (as shown in Figs.~\ref{Figure3}(a) and \ref{Figure3}(b)), is explained by an atom located away from the centre closer to the left gate in the experiment. The overall pumping behaviour is constant for displacements of 15 nm both towards the left and right gate, resulting in a robust single-atom electron pump over a distance of 30 nm (see Fig.~\ref{Figure5}(d)). This is a crucial observation for establishing a reproducible charge pump geometry. The current single atom ion implantation techniques have an accuracy around 15 nm~\cite{Alves:2013aa}, which could enable a controllable fabrication of accurate single-atom electron pumps. From Fig.~\ref{Figure5}(d) it is also concluded that moving the atom further to the right gate significantly improves the accuracy of the single-atom electron pump. The improvement in accuracy is a consequence of the weaker coupling of the atom to the left gate, which as a result keeps the atomic states longer below the Fermi energy ($\lvert t_3-t_2 \rvert$ is increased) and therefore has the same effect as a decrease in pumping frequency. However, this high accuracy region is less robust to displacements of the atomic potential. 

\section{Conclusion}
In conclusion, a model that describes the behaviour of a single parameter single-atom electron pump has been presented. This model describes the loading, isolating and unloading of the electron via the robust Coulomb potential of the atom as the main steps for each pumping cycle. The most striking feature if comparing with dynamical quantum dot pumps, is the fact that, for atom pumps, excited states greatly enhance the accuracy by increasing the loading efficiency of the ground state via a fast relaxation process. This is in agreement with what has been recently observed experimentally by several groups~\cite{Tettamanzi:2014aa, Yamahata:2014aa} and observed in the experiment presented here. The model allows to benchmark against the position of the atom in the channel of the silicon transistor and shows that the single-atom pumps performances are unaffected by displacements of the atom up to a few tens of nm, which are at reach of the precision achieved by the current atom placement technologies~\cite{Alves:2013aa}. Lastly, the accuracy of these pumps could be enhanced by working with more effective pulse shapes and by increasing the relaxation rate to the ground state of the atom, making single-atom electron pumps an even more attractive alternative to dynamical quantum dot charge pumps.

%%%%%%%%%%%%%%%   ACKNOWLEDGMENTS  %%%%%%%%%%%%%%%
\section*{Acknowledgments} 
The authors gratefully acknowledge J. Verduijn for developing a finite difference Poisson solver and L. Fricke for fruitful discussions. The authors acknowledge financial support from DP120101825. G. C. Tettamanzi acknowledges financial support from the ARC-DECRA scheme DE120100702, project title : ’Single Atom Based Quantum Metrology’.

%%%%%%%%%%%%%%%   BIBLIOGRAPHY  %%%%%%%%%%%%%%%
%


\begin{thebibliography}{40}
\bibitem [{not()}]{note1}%
  \BibitemOpen
  \href@noop {} {}\bibinfo {note} {An estimation of the attempt frequency for
  the donor is made by imagining an electron bouncing back and forth between
  two walls and counting the frequency with which it hits one of these walls.
  To get an estimation of the 'speed' we use the binding energy as a kinetic
  energy and the Bohr radius is used as the distance between the walls. This
  leads to a $\Gamma_0$ of $2 E_b / \hbar$, which is close to 100 THz for the
  Phosphorous ground state.}\BibitemShut {Stop}%
\end{thebibliography}
\end{document}